\documentclass[letterpaper, 10 pt, conference]{ieeeconf}  

\IEEEoverridecommandlockouts                              

\usepackage{graphicx} 
\usepackage{caption}
\usepackage{subcaption}

\usepackage{amsmath}
\usepackage{cite}
\usepackage[outdir=./]{epstopdf}
\usepackage[font=small]{caption}
\title{\LARGE \bf
Covariance intersection to improve the robustness of the photoplethysmogram derived respiratory rate}

\author{Jia Zhang, Gaetano Scebba and Walter Karlen, \it{Senior Member}, IEEE
\thanks{The work was supported by the Swiss National Science Foundation SNSF (150640).}
\thanks{Mobile Health Systems Lab, Institute of Robotics and Intelligent Systems, Department of Health Sciences and Technology, ETH Zurich, Zurich, Switzerland (email:
		{\tt\small jia.zhang@hest.ethz.ch},
        {\tt\small walter.karlen@ieee.org})}%
}

\begin{document}

\maketitle
\thispagestyle{empty}
\pagestyle{empty}

\begin{abstract}
Respiratory rate (RR) can be estimated from the photoplethysmogram (PPG) recorded by optical sensors in wearable devices. The fusion of estimates from different PPG features has lead to an increase in accuracy, but also reduced the numbers of available final estimates due to discarding of unreliable data. We propose a novel, tunable fusion algorithm using covariance intersection to estimate the RR from PPG (CIF). The algorithm is adaptive to the number of available feature estimates and takes each estimates' trustworthiness into account. In a benchmarking experiment using the CapnoBase dataset with reference RR from capnography, we compared the CIF  against the state-of-the-art Smart Fusion (SF) algorithm. The median root mean square error was 1.4 breaths/min for the CIF and 1.8 breaths/min for the SF. The CIF significantly increased the retention rate distribution of all recordings from 0.46 to 0.90 (p~$<$~0.001). The agreement with the reference RR was high with a Pearson's correlation coefficient of 0.94, a bias of 0.3 breaths/min, and limits of agreement of -4.6 and 5.2 breaths/min. In addition, the algorithm was computationally efficient. Therefore, CIF could contribute to a more robust RR estimation from  wearable PPG recordings.

\end{abstract}

\section{INTRODUCTION}


Respiratory rate (RR) is an essential vital sign to assess the medical condition of patients and abnormal RR is an important predictor of serious illness \cite{Jayaraman2008}. Continuous monitoring of RR and RR trend changes can detect abnormal events among general ward patients and enable early interventions \cite{Mok2015}. 
Commonly used sensing methods in the clinic, such as capnometry and spirometry, are based on the analysis of ex- and inhaled air (i.e. gas flow and composition changes), but are cumbersome to wear, obstructive, or subject to strong artefacts if the environment is not controlled. Wearable devices show great potential for unobstructive and continuous monitoring and could overcome some of these limitations. 

Objectively assessed RR with mobile sensors shows potential for improving the diagnosis of acute lower respiratory infections at the point-of-care \cite{Karlen2014}. RR can be estimated by analyzing the photoplethysmogram (PPG), which is increasingly available on wearable devices. The PPG waveform is known to have multiple modulations induced by respiration, such as respiratory-induced intensity (RIIV), amplitude (RIAV), frequency (RIFV) \cite{Karlen2013c}, width (RIWV)\cite{Lazaro2013b}, and slope transit time variation (RISV) \cite{Addison2016c}. The current most cited benchmark method for RR estimation from the PPG signal is Smart Fusion (SF) \cite{Karlen2013c}. It fuses three respiratory-induced variations (RIIV, RIAV, RIFV) by calculating their mean and discards RR estimates that are labelled as artefact or where the standard deviation between the three estimates is larger than $4$ breaths/min. 

Research for improving the accuracy of RR estimation algorithms has primarily focused on discarding poor quality data points and did put less efforts in retaining or reconstructing information. For example, the SF discards a high percentage of estimates in order to retain only accurate RR information \cite{Karlen2013c}. However, even if the PPG signal is shown to be of good quality, the RR is not computed due to the disagreements between different RR estimates. Pimentel et al. show that on average, $36\%$ of data is eliminated in SF due to such disagreements \cite{Pimentel2016}. Another estimation approach is to apply Kalman filters to each individual estimate leading to smoother time series for RR estimation \cite{Nemati2010c}. However, this approach dampens rapid changes in RR and therefore reduces the detection of abnormal RR events that  would be clinically relevant. 

Abnormal event detection relies on robust and continuous RR measurements \cite{Birrenkott2017b}. Robust measurements are defined by high accuracy of the estimation and low missing data, i.e. high retention rates after the fusion. It is crucial to retain as much continuous information as possible because data gaps might lead to missed signs of abnormal events and cause failure in recognizing patient deterioration \cite{ Mok2015,Prgomet2016}.

We present a novel fusion approach to extract RR from the PPG signal. Our aim was to develop an algorithm that 1) provides robust RR estimations independently of the quality of individual estimates, and 2)  can be simultanously optimized for clinical required accuracy and  high retention rates. We characterized this algorithm against a benchmark dataset and the established SF algorithm. 
\section{Methods}
\begin{figure*}[!htb]
\centering
\includegraphics[width=1\textwidth]{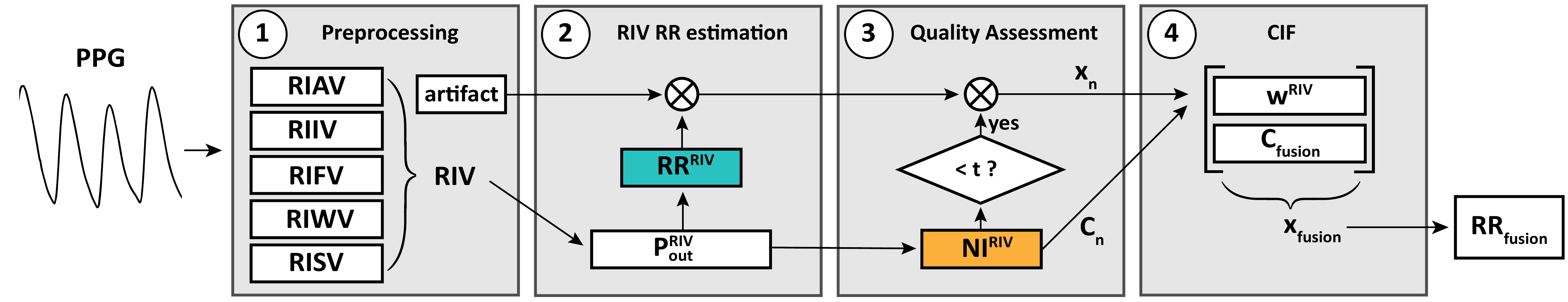}
\caption{RR covariance intersection fusion (CIF): The PPG waveform was preprocessed and five different respiratory-induced variations (RIV)  and artifacts were extracted (1). Spectral analysis was conducted for each RIV to subsequently estimate the respiratory rate of each variation (RR$^{RIV}$, 2) and a noise index (NI$^{RIV}$, 3). The quality of each estimation was assessed by comparing NI to a threshold t. CIF fused all good quality RR estimates $x_n$ using the NI derived covariance $C_n$ (4) to compute a fused RR estimation (RR$_{fusion}$).}
\label{fig_fusionflowchart}
\end{figure*}

We proposed an algorithm that integrates covariance intersection fusion (CIF) as a central element to join the RRs extracted from the available PPG features (Fig. \ref{fig_fusionflowchart}): 1) In a preprocessing step, the algorithm extracted the five RR induced variations from the PPG signal (RIAV, RIIV, RIFV, RISV, RIWV). 2) The dominant frequency for each variation was extracted from the maximum spectral power peak obtained by the FFT and a subtracted fitted power law function. 3) The quality of each estimation was determined by considering the noise index (NI) of each variation and an estimation was eliminated when this NI was too low. 4) Finally, the remaining RR estimates were fused into a final RR$_{fusion}$ using the CIF method.

\subsection{Preprocessing}
We adopted the preprocessing steps from the SF approach that included filtering, pulse segmentation and artifact detection \cite{Karlen2013c}. The RIAV, RIIV and RIFV were extracted using the same methods as described in \cite{Karlen2013c}. The RIWV and RISV were extracted using the algorithms presented in \cite{Lazaro2013b} and \cite{Addison2016c}. Each RR induced variation (RIV) signal was re-sampled to $5$ Hz.  

\subsection{Respiratory induced variation RR estimation}
Analogous to the procedure in SF algorithm, when no artifact was detected we used the FFT to calculate the maximum spectral component within the RR band. The FFT window was 32 s with a shift of 2 s. We adapted the method by additionally introducing a power law fitting to reduce background noise \cite{Gottselig2002}. First, we fitted a linear function into the log-log scale of the spectrum in the range of 2 to 4 breaths/min and 65 to 100 breaths/min, excluding the range of interest between 4 to 65 breaths/min such as
\begin{equation}
log(P_{\text{fit}}) = a \times log(f) + k,
\label{equ_logfit}
\end{equation}
\begin{equation}
P_{\text{fit}} = \exp^{k} \times f^{a},
\label{equ_expfit}
\end{equation}
  where $P_{\text{fit}}$ was the fitted power spectrum, $f$ was the frequency, $a$ was the slope of the fitting line and $k$ was the intersection.  Very low ($<$2 breaths/min) and very high RR ($>$100 breaths/min) were not considered for the fitting because of possible ambiguous artefacts in these RR ranges. We then subtracted the fitted values from the power spectrum $P$ such as 
\begin{equation}
P_{out} = P - P_{\text{fit}}.
\label{equ_outP}
\end{equation}
We detected the maximum peak $P_{\text{out}_{\text{peak}}}$ of the subtracted spectrum $P_{\text{out}}$ within the RR range of interest ($ \:j \in 4-65 \:$ breaths/min) and its corresponding RR to be the estimate for each RR variation. 

\subsection{Estimate quality assessment}

To evaluate the quality of each estimate, we calculated the NI such as
\begin{equation}
NI  = \dfrac{P_{\text{out}_{\text{peak}}}}{\sum_{j} P_{\text{out}_j}}. 
\label{equ_noiseratio}
\end{equation}
The NI described how discriminative the primary detected RR was from all other possible RR with a range from 0 to 1.
If the NI for a RR estimate was smaller than a predefined threshold $t$, that estimate was not forwarded to the CIF. 
 
\subsection{Covariance intersection and RR fusion} 

CIF is an algorithm for fusing estimates with unknown correlation that origins from the field of control and estimation \cite{Julier2009}. It considers the error correlation between the sources of estimates, which can achieve robustness and consistency in data fusion \cite{Julier2009}. CIF takes the convex combination of the mean and covariance of the estimates expressed in information space. For an arbitrary number n $>$ 2 of estimates ($x_1, x_2...x_n$), which are corrupted by measurement noises and modelling errors, the estimates can be fused into a final estimation $x_{\text{fusion}}$. The intersection is the convex combination of the covariances $C_{1}, ...C_{n}$, such as
\begin{align}
C_{\text{fusion}}^{-1} &= \omega_1C_{1}^{-1} + ... + \omega_nC_{n}^{-1}, \\
C_{\text{fusion}}^{-1}x_{\text{fusion}} &= \omega_1C_{1}^{-1}x_1 + ... + \omega_nC_{n}^{-1}x_n,  \\
&\sum_{i=1}^{n} \omega_i = 1. \label{equ_weights}
\end{align}
where $ \omega_n $ was the weight of the estimate. The CIF algorithm estimated RR from multiple different RIVs obtained from a single PPG signal with unknown correlations. 
Therefore, the uncertainty of each estimate described by the NI was used as covariance term for the CIF, such as
\begin{equation} 
C_{n}= 1-\text{NI}_{n}.
\end{equation} 

\subsection{Computational efficiency}
We designed the CIF as a one-dimensional function. 
The condition 
\begin{equation}
\omega_1C_{1} = \omega_2C_{2} = ... = \omega_nC_{n}
\end{equation}
solved eq.~\ref{equ_weights} for $\omega_n$ 
using $n$ equations. This kept the computational effort with $n$ division or multiplications relatively low compared to the previous processing steps.  
 The entire algorithm was implemented in Matlab (R2019b, MathWorks Inc, Natick, USA).  

\subsection{Dataset}

We tested our algorithm on the TBME RR benchmark dataset from CapnoBase \cite{Karlen2010f}. The benchmark dataset contained 42 8-minute PPG recordings from 29 pediatric and 13 adult subjects undergoing elective surgery with either spontaneous or controlled ventilation. For comparison, the manually annotated data from capnography served as reference RR.

\subsection{Evaluation}

 For a more objective comparison, we computed the SF using all the above mentioned five RIV estimates (SF5) in addition to the three original benchmark estimates (SF3) \cite{Karlen2013c}. We evaluated the root mean squared error (RMSE) breath-by-breath across subjects for the fused RR against the reference RR from the benchmark dataset. We defined a retention rate as the ratio of the number of retained RR windows to the total possible number of RR windows within a recording. We varied the NI threshold $t$ between 0 and 0.3 with an increment of 0.01 to depict the relationship between retention rate and RMSE. 
 
 To highlight the differences between the algorithms, we fixed t to 0.13 
 and calculated retention rate and RMSE for each subject. The results were depicted with boxplots. We conducted the Wilcoxon test with Bonferroni correction to compare the RMSE and retention rate distributions obtained by each algorithm. Finally, with computing Bland-Altman statistics and the Pearson's correlation coefficient $r$  we investigated the agreement between the proposed CIF and the reference RR. 
   
\section{Results}
\begin{figure}[!t]
	\centering
	\includegraphics[width=.50\textwidth]{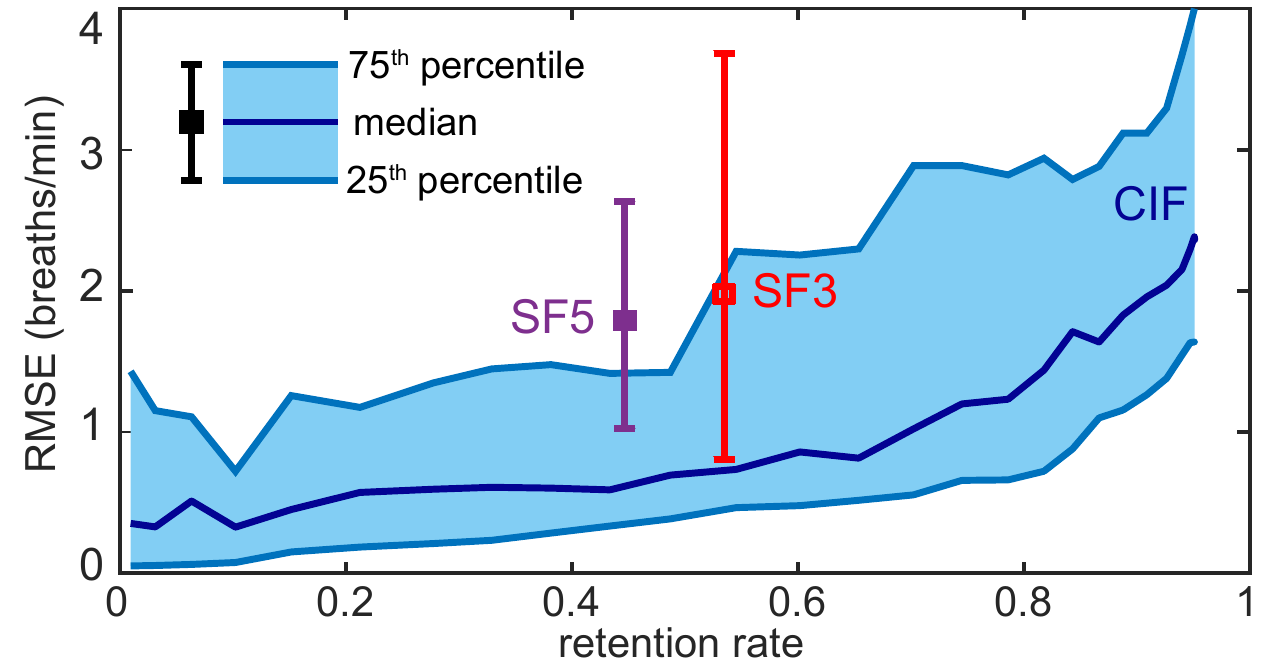}
	\caption{Retention rate vs. RMSE of the proposed CIF otained by varying the noise index threshold from 0 to 0.3  (blue shaded area). The RMSE increased with an increasing  retention rate. The middle, lower and upper bound lines were median, 25$^{th}$ and 75$^{th}$ percentile of the RMSE for CIF, SF3, and SF5.  }
	\label{fig_rms_retention}
\end{figure}

\begin{figure}[!t]
\centering
\includegraphics[width=0.49\textwidth]{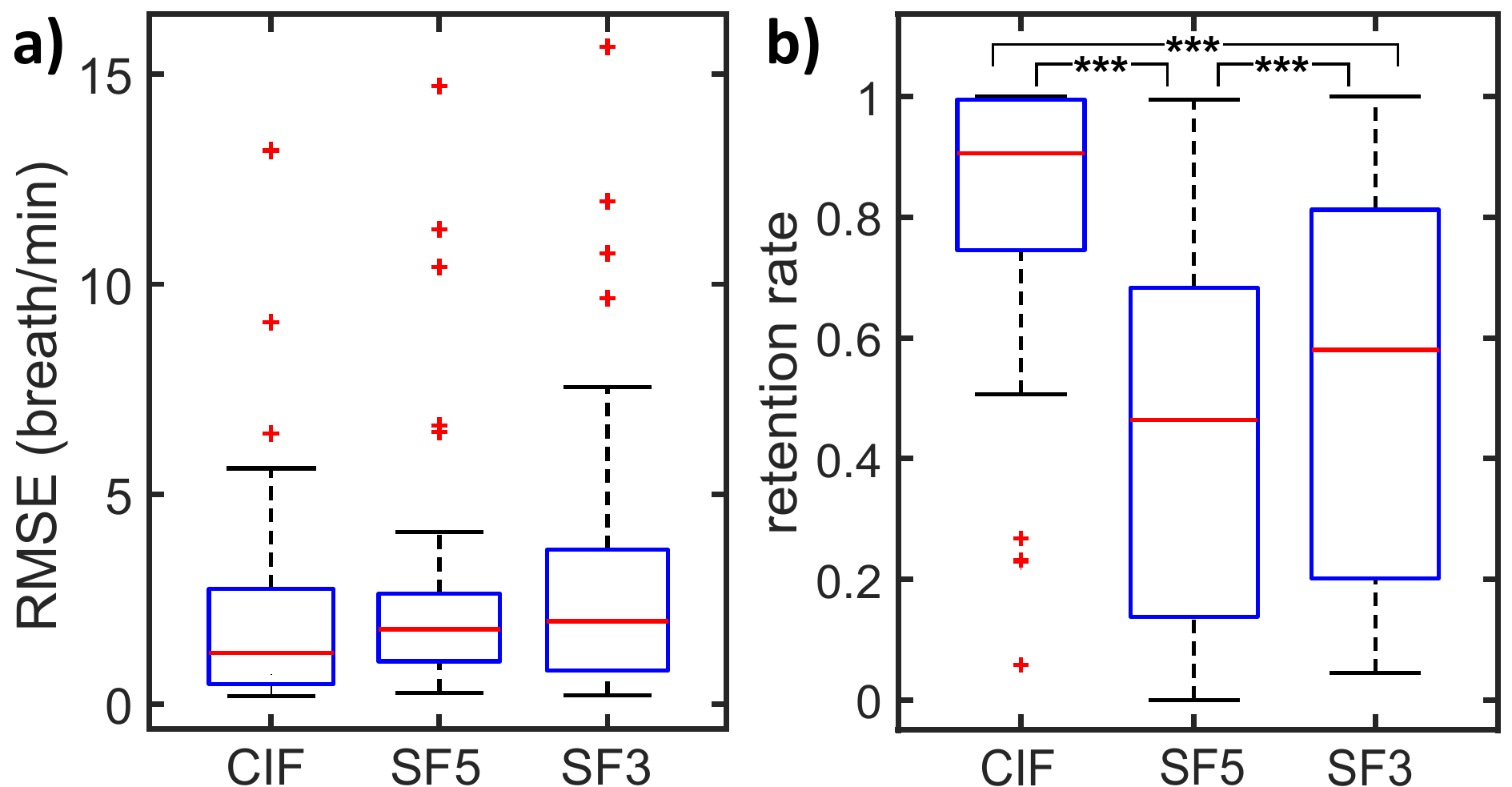}
\caption{Capnobase Benchmark comparison of CIF, SF with five estimates (SF5) and three estimates (SF3) for: a) RMSE and b) retention rate across subjects. The central, bottom and top edge of the box indicated the median, 25$^{th}$ and 75$^{th}$ percentiles. The outliers were depicted as a cross. The 3 asterisks indicated the significance level of the distribution difference (*** p$<$0.001). }
\label{fig_rms_retention_boxplot}
\vspace{0.5cm}
\end{figure}

\begin{figure}[!t]
	\centering
	\includegraphics[width = .49\textwidth]{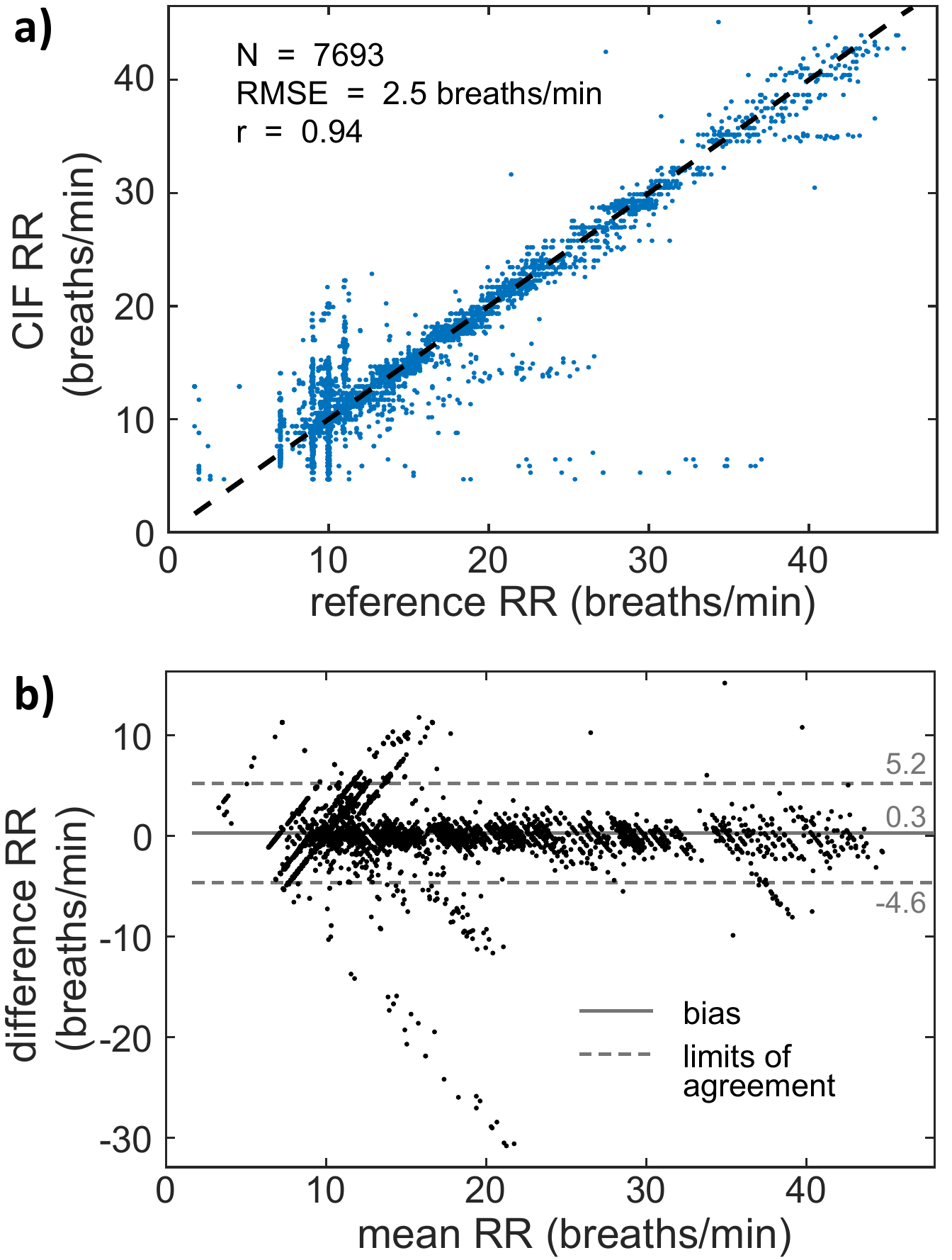}
	\caption{Agreement between the CIF fused and the reference RR on the CapnoBase benchmark. a) scatter plot; b) Bland-Altman plot with the middle line representing the mean error (bias) and the two dash lines representing the limits of agreement.}
	\label{fig_BASC}
\end{figure}

The RMSE for the CIF increased when the retention rate increased (Fig. \ref{fig_rms_retention}). The SF5 showed lower retention rate than SF3 (Fig. \ref{fig_rms_retention}). The proposed CIF showed lower RMSEs compared to the SF when keeping the retention rate fixed. 
The median RMSE over all 42 subjects was 1.4 breaths/min for the CIF compared to 1.8 breaths/min for the SF5 and 2.0 breaths/min for the SF3 (Fig. \ref{fig_rms_retention_boxplot}$a$). The CIF significantly increased the overall retention rate distribution from 0.46 (SF5) and 0.58 (SF3) to 0.90 (p~$<$~0.001, Fig. \ref{fig_rms_retention_boxplot}b).  
Compared to the reference RR,  $r$ was 0.94 (Fig. \ref{fig_BASC}a), the bias was 0.3 breaths/min and the limits of agreement were -4.6 and 5.2 breaths/min (Fig. \ref{fig_BASC}$b$). 

\section{Discussion}
We developed and  evaluated a novel fusion algorithm to estimate RR from PPG signal based on five RR induced variations. The algorithm showed improved retention rates and comparable high accuracy  to the benchmark algorithm. The CIF based algorithm was adaptive and could fuse an arbitrary number of RR estimates independently of missing estimates that were eliminated during the quality control process. A simple parameter enabled the tuning of the performance. 

The NI threshold offered an effective way to tune the algorithm and optimize the RMSE and retention rate. There was an unavoidable trade-off between RMSE and retention rate. This method would enable a transparent way to align an application dependent task with a desired performance for the algorithm. 

We evaluated the accuracy and retention rate of SF by fusing all five estimates and the three estimates used in \cite{Karlen2013c}. Fusing all five estimates showed decreased median RMSE, but also decreased retention rate. This indicated that the two additional variations RIWV and RISV added value to the fusion. However, more estimates also introduced a higher degree of uncertainty that lead to less estimates. This indicated that considering only the agreement among estimates was not sufficient to extract a robust RR. More estimates were leading to higher likelihoods for the precence of a disagreement, leading to a lower retention rate. Under the consideration that all the RR estimates were extracted from a single PPG signal, causal dependencies among the estimates might have existed. Our algorithm did not treat the estimates independently, but considered the potential correlation and redundancy among estimates to reduce the fusion error. Therefore, our proposed method was more robust in terms of accuracy and retention rate compared to the SF algorithms.   

The proposed CIF algorithm was computationally efficient. The average processing time for 30 RR estimations was 0.49 s on a single core CPU 2.60 GHz PC. Other approaches that focused on optimizing retention rate are computationally much more costly. For example, multiple steps of time-frequency dependencies (FFT, auto-correlation and auto-regression) need to be computed in \cite{Birrenkott2017b}. For embedded applications power efficiency is key. Therefore, our approach offers a clear advantage  for real-time assessments on wearables with limited computational power.

The CIF algorithm was evaluated only on one benchmark dataset that was limited in size and population. Evaluation on larger datasets including data that was acquired with wearable devices will be needed to better understand the performance of the CIF with altered sources of noise. For implementation, we recommend a prospective study with extensive validation experiments.

RR is one of the important vital signs that provides objective and convincing medical evidence to describe a patient's health status, yet it is often neglected \cite{Jayaraman2008, Mok2015}. When continuous monitoring of RR estimation is essential for identifying abnormal events, low RR retention rates might lead to events missed. 
By simply considering the agreement among estimates, the current benchmark fusion method can easily fail when individual respiratory-induced variations are not prominent. This can be seen in in specific diseases or populations, such as in critically ill or elderly where the  absence of modulation from the autonomic system decreases or cancels the RIFV \cite{Meredith2012}. The main contribution of our proposed algorithm is that it can estimate the RR independently of the number of RIVs available. The decision to take a variation into consideration was taken at each fusion independently, making the system adaptive to changes in the input conditions.
\newpage
 For remote assessment applications, where medical experts are not available to operate the devices and interpret the measured signal, it is essential that the collected data are of high integrity \cite{Zhang2018b}. In this work, we focused on improving two aspects of data integrity, i.e. accuracy (performance) and retention rate (completeness), and provided a tuning mechanism to optimize between the two. This approach is of high interest for robust RR estimation in wearable devices that measure PPG with optical sensors.
 

\addtolength{\textheight}{-12cm}   




\bibliographystyle{IEEEtran}
\bibliography{MHealth-RRfusion}

\end{document}